\documentclass[11pt, a4paper, oneside, reqno]{amsart}
\usepackage{comment}
\usepackage[usenames, dvipsnames]{color}
\definecolor{darkblue}{rgb}{0.0, 0.0, 0.45}
\definecolor{lightblue}{RGB}{240,248,255}
\definecolor{lightblue2}{rgb}{0.68, 0.85, 0.9}
\definecolor{lightcyan}{rgb}{0.88, 1.0, 1.0}
\definecolor{palepink}{rgb}{0.98, 0.85, 0.87}

\usepackage[colorlinks	= true,
raiselinks	= true,
linkcolor	= darkblue, 
citecolor	= Mahogany,
urlcolor	= ForestGreen,
pdfauthor	= {Peyman Mohajerin Esfahani},
pdftitle	= {},
pdfkeywords	= {},
pdfsubject	= {},
plainpages	= false]{hyperref}

\usepackage{dsfont,amssymb,amsmath,enumitem} 
\usepackage{amsfonts,dsfont,mathtools, mathrsfs,amsthm} 
\usepackage[amssymb, thickqspace]{SIunits}
\usepackage{fancyhdr,mdframed,nicefrac}

\usepackage[norelsize, ruled, vlined]{algorithm2e}
\usepackage{algcompatible}

\usepackage{epsfig}
\usepackage{graphicx}
\usepackage{float}
\usepackage{caption}
\usepackage{subcaption}

\usepackage{courier}

\usepackage{multirow}
\usepackage{bigstrut}

\allowdisplaybreaks
\date{\today}
\makeatletter
\def\@settitle{\begin{center}%
		\baselineskip14\p@\relax
		\normalfont\LARGE\scshape\bfseries
		\@title
	\end{center}%
}

\def\@setauthors{%
  \begingroup
  \def\thanks{\protect\thanks@warning}%
  \trivlist
  \centering\footnotesize \@topsep30\p@\relax
  \advance\@topsep by -\baselineskip
  \item\relax
  \author@andify\authors
  \def\\{\protect\linebreak}%
  \authors%
  \ifx\@empty\contribs
  \else
    ,\penalty-3 \space \@setcontribs
    \@closetoccontribs
  \fi
  \endtrivlist
  \endgroup
}

\makeatother
\makeatletter

\def\subsection{\@startsection{subsection}{2}%
	\z@{.5\linespacing\@plus.7\linespacing}{.5\linespacing}%
	{\normalfont\large\bfseries}}

\def\subsubsection{\@startsection{subsubsection}{3}%
	\z@{.5\linespacing\@plus.7\linespacing}{.5\linespacing}%
	{\normalfont\itshape}}

\usepackage{multirow}
\usepackage{bigstrut}
\usepackage{wrapfig}
\usepackage{caption}

\renewcommand{\geq}{\geqslant}
\renewcommand{\ge}{\geqslant}

\renewcommand{\leq}{\leqslant}

\DeclareSymbolFont{symbolsC}{U}{pxsyc}{m}{n}

\usepackage{bm}
\usepackage{amsthm}

\usepackage{amssymb}
\usepackage{latexsym}
\usepackage{mathrsfs}  
\usepackage{bbm}
\usepackage{amsthm}

\usepackage{url}
\usepackage{color}

\usepackage{algorithm2e}
\usepackage{amsmath,bm}
\usepackage{subcaption}
\usepackage[makeroom]{cancel}

\usepackage{float}
\usepackage{lipsum}
\usepackage{dsfont}
\usepackage{rotating}
\usepackage{mathtools}
\usepackage{relsize}

\definecolor{referee1}{RGB}{0, 0, 0}
\definecolor{referee2}{RGB}{0, 0, 0}
\definecolor{mine}{RGB}{0, 0, 0}
\definecolor{referee1_2nd}{RGB}{0,0,0}
\title[Enskog-Vlasov description of dense fluids]{
Treatment of long-range interactions arising in the Enskog-Vlasov description of dense fluids
}
 \thanks{Corresponding author: Mohsen Sadr}
 \thanks {Email: sadr@mathcces.rwth-aachen.de}
\thanks{Mohsen Sadr: MATHCCES, Department of Mathematics, RWTH Aachen University, Schinkestrasse 2, D-52062 Aachen,
Germany. M. Hossein Gorji: MCSS, Ecole Polytechnique F{\'e}d{\'e}rale de Lausanne (EPFL), CH-1015 Lausanne, Switzerland.}
\date{February 1, 2019}

\begin{document}
\maketitle

\begin{abstract}
The kinetic theory of rarefied gases and numerical schemes based on the Boltzmann equation, have evolved to the cornerstone of non-equilibrium gas dynamics. However, their counterparts in the dense regime remain rather exotic for practical non-continuum scenarios. This problem is partly due to the fact that  long-range interactions arising from the attractive tail of molecular potentials, lead to a computationally demanding Vlasov integral. This study focuses on numerical remedies for efficient stochastic particle simulations based on the Enskog-Vlasov kinetic equation. In particular, we devise a Poisson type elliptic equation which governs the underlying long-range interactions. The idea comes through fitting a Green function to the molecular potential, and hence deriving an elliptic equation for the associated fundamental solution. Through this transformation of the Vlasov integral, efficient Poisson type solvers can be readily employed in order to compute the mean field forces. Besides the technical aspects of different numerical schemes for treatment of the Vlasov integral, simulation results for evaporation of a liquid slab into the vacuum are presented. It is shown that the proposed formulation leads to accurate predictions with a reasonable computational cost. 
\end{abstract}


\section{Introduction}
\label{sec:introduction}
\noindent It is well known that non-equilibrium phenomena which appear around the critical point condition may not be captured accurately via conventional hydrodynamics \cite{frezzotti2005mean,ishiyama2004molecular}. The situation is imminent in the phase transition phenomena arising e.g. in 
  fuel droplets \cite{sazhin2017modelling}, molecular distillation \cite{lutivsan1995mean,wang2009separation,li2014dsmc}, evaporation processes in the laser-solid interaction \cite{knight1979theoretical}, sonoluminescence \cite{brenner2002single} and supercritical gas-liquid interface in injecting engines \cite{DAHMS20131667,Dahms20153648,dahms2015non}. 
 The kinetic theory provides an attractive framework which overcomes the closure problem associated with macroscopic quantities, through notion of the distribution function.  At the same time, the computational complexity is still significantly less than molecular dynamics (MD) based simulations \cite{frezzotti2017kinetic}. \\ \ \\
\noindent The classical kinetic theory leads to a hierarchy of evolution equations for the single molecular distribution. In particular, it provides the Boltzmann equation for the ideally dilute gas state, the Enskog equation for the dense gas regime (in the absence of long-range interactions) and finally the Enskog-Vlasov equation for the dense fluids subject to the mean-field interactions \cite{karkheck1981kinetic, grmela1971kinetic}. \\ \ \\
Three categorically different methodologies have been devised in order to analyze and simulate flow phenomena based on the kinetic models. First category belongs to the moment methods, where the governing equation is projected to a finite set of moments  (see e.g. \cite{torrilhon2016modeling, kremer1988enskog}). A high fidelity approach can be constructed by the so-called direct (kinetic) schemes, where the probability density is discretized in the whole phase space (see e.g. \cite{broadwell_1964}).
{\color{referee1_2nd}
Accordingly, spectral methods have been developed for computing the collision operator 
\cite{gamba2009spectral,wu2013deterministic,wu2015fast,wu2016non}.}
Finally, a commonly used approach is based on the stochastic particle methods, where computational particles are employed as Monte-Carlo samples of the distribution (pioneered by Bird \cite{Bird1963}). The latter leads to the Direct Simulation Monte-Carlo (DSMC) for rarefied gas simulations based on the Boltzmann equation \cite{Bird}. \\ \ \\
The DSMC counterparts for dense gases include Enskog Simulation Monte-Carlo (ESMC) \cite{montanero1997simulation,montanero1997viscometric,Montanero1996}, {\color{referee2} Frezzotti}'s algorithm \cite{Frezzotti1997} and consistent Boltzmann algorithm (CBA) \cite{Alexander1995}. While ESMC and Frezzotti's algorithms have been shown to accurately solve the Enskog equation, their computational cost increases with the density. This is due to the fact that the collision rate scales with the density squared which leads to a significant number of jumps implied by the Markov process underlying the Enskog collision operator. As an efficient alternative, following the methodology proposed by Jenny \textit{et al.} \cite{Jenny2010} and Gorji \textit{et al.} \cite{Gorji2011,Gorji2014} for ideal gases, a Fokker-Planck model as an approximation of the Enskog operator was introduced by the authors \cite{sadr2017continuous}. In the Fokker-Planck approach, the effect of collisions are modeled through drift and diffusion actions which lead to continuous stochastic paths instead of binary collisions. \\ \ \\
Apart from the dense computations arising from binary collisions, the long-range interactions impose yet another computational challenge for flow simulations based on the kinetic theory.
 Note that even though the Enskog equation has shown reasonable consistency with MD simulations for dense gases, the attractive part of the inter-molecular potential can no longer be ignored once liquid flows or the phase transition are encountered. In order to cope with these interactions, the Vlasov mean-field limit \cite{vlasov1978many} has been carried out, leading to the Enskog-Vlasov equation. The  resulting Vlasov integral then can be considered as a macro-scale inter-molecular potential which appears as convolution of the number density and the mean-field potential in the physical space.
 \\ \ \\
\noindent In order to compute the mean-field forces modeled based on the Vlasov integral, so far mainly two approaches have been devised and discussed. First approach relies on a direct computation of the integral e.g. by using a particle representation of the distribution \cite{frezzotti2005mean,piechor1994discrete,frezzotti2018mean}. While here, high fidelity computations can be performed, the scheme results in an N-body type complexity. A computationally more efficient method is obtained by considering the Taylor expansion of the number density with the assumption that its higher order derivatives become zero \cite{he2002thermodynamic}. Unfortunately this assumption yields a strong constraint on the number density which may not be universally justified. \\ \ \\
Even though the computational complexity resulting from the attractive forces does not scale with the density,  it can become the bottle-neck of multi-phase flow simulations once the collision operator is replaced by an efficient continuous stochastic model, such as the Fokker-Planck model \cite{sadr2017continuous}.
The main objective of this study is to introduce a numerical scheme for solving the Vlasov integral, where the resulting computational complexity is less demanding than the direct method. At the same time, the proposed scheme does not suffer from strong assumptions underlying the density expansion method. This is achieved by transforming the Vlasov integral to a Poisson type equation. Note that similar ideas have been already employed in plasma community where the Coulomb interactions are treated by the Poisson equation (see e.g. \cite{thomas2012}).  Once the integral is transformed to the corresponding elliptic problem, available numerical schemes are adopted for efficient simulations of flows subject to long-range interactions. \\ \ \\ 
\noindent The rest of the paper is organized as the following. First in \S~\ref{sec:kinetic_model}, we provide a review of the kinetic theory including some aspect of the Enskog-Vlasov equation. 
Next, current approaches in computing the Vlasov integral i.e. direct and density expansion methods are reviewed in \S~\ref{sec:treatment}. Then, the main idea of the paper is devised, where the solution of the Vlasov integral is linked to the screened-Poisson equation for the Sutherland molecular potential. 
In \S~\ref{sec:results}, performance and accuracy of the devised scheme is investigated through numerical experiments. 
The paper is concluded in \S~\ref{sec:conclusion} with a short overview of the outlooks.

\section{Review of the Enskog-Vlasov formalism}
\label{sec:kinetic_model}
\noindent Consider an ensemble of neutral monatomic particles, each of which has the mass $m$ and is subject to the Sutherland molecular potential \cite{Hirschfelder1963,sutherland1893lii}
\begin{eqnarray}
\phi(r)=
\left\{\begin{array}{lr}
 +\infty & r<\sigma,\\ 
 -\phi_0 \Big (\dfrac{r}{\sigma} \Big )^{-6}& r \geq \sigma
\end{array}\right. 
\label{eq:sutherland}
\end{eqnarray}
where $\bm r:= \bm y - \bm x$ denotes the relative position of a particle at $\bm x$ with respect to the one at $\bm y$. 
Note $r := |\bm r|$  where here and henceforth $|\ . \ |$ indicates the usual Euclidean norm. The coefficient $\phi_0$ and the effective diameter $\sigma$ are determined empirically and have  specific values for a given system of identical monatomic molecules.
Let $f(\bm v; \bm x, t)$ denote the density associated with the probability of finding a particle with the velocity close to $\bm v$, at a given  position $\bm x$ and an instant in time $t$.  For convenience, let $\mathcal{F}(\bm v, \bm x, t) := \rho(\bm x, t) f(\bm v; \bm x, t)$ denote the mass distribution function (MDF), hence
\begin{flalign}
\rho(\bm x, t) = \int_{\mathbb{R}^3} \mathcal{F} d^3 \bm v~,
\end{flalign}
where $\rho=nm$ is the density, $n$ reads the number density{\color{referee2}, $\mathcal{F}$ is short for $\mathcal{F}(\bm v, \bm x, t)$} and $d^p \bm l = dl_1...dl_p$. 
\subsection{The Enskog-Vlasov Equation}
\noindent The main objective of the kinetic theory is to provide an evolution equation governing the dynamics of $\mathcal{F}$. Yet since $\mathcal{F}$ stands for a single point velocity distribution, further assumptions are required in order to simplify the multi-point correlations. 	A closed form equation can be obtained by neglecting the correlations in the soft-tail of the potential i.e. $\mathcal{F}(\bm{v_1}, \bm{x_1}, \bm{v_2}, \bm{x_2},t) = \mathcal{F}(\bm{v_1}, \bm{x_1},t) \mathcal{F}(\bm{v_2}, \bm{x_2},t)$ for $r>\sigma$, while including the contact value of the pair correlation function $Y$  in the collision operator i.e. $\mathcal{F}(\bm{v_1}, \bm{x_1}, \bm{v_2}, \bm{x_2},t) = Y((\bm x_1+\bm x_2)/2)\mathcal{F}(\bm{v_1}, \bm{x_1},t) \mathcal{F}(\bm{v_2}, \bm{x_2},t)$ for $r<\sigma$. 
Furthermore, by adopting a force field $\bm \xi(\bm x,t)$ as the mean-field limit of the long-range interactions along with the Enskog collision operator $S^{\text{Ensk}}(\mathcal{F})$ for the short-range encounters, the Enskog-Vlasov equation can be derived \cite{karkheck1981kinetic,grmela1971kinetic}, leading to 
\begin{flalign}
\frac{\partial \mathcal{F}}{\partial t}
+ \frac{\partial (\mathcal{F} v_i)}{\partial x_i}
- \frac{\xi_i}{m} \frac{\partial \mathcal{F}}{\partial v_i} 
= S^{\text{Ensk}}(\mathcal{F})
\label{eq:full_vlasov_enskog_eq}
\end{flalign}
for our setup of monatomic molecules with the Sutherland potential \cite{frezzotti2005mean}. 
 Here, the attractive force 
\begin{flalign}
\xi_i(\bm x, t) &= \frac{ \partial U}{\partial x_i} \hspace{0.5cm} 
\end{flalign}
 is gradient of the mean-field potential 
\begin{eqnarray}
U(\bm x, t)   &=&  \int_{r>\sigma}\phi( r) n(\bm y, t) d^3 \bm y~,
\label{eq:vlasov_integral}
\end{eqnarray}
known as the Vlasov integral. Notice that the Einstein summation convention is employed throughout this work.  \\ \ \\
Since no assumption has been made with regard to the phase of the matter, the kinetic model \eqref{eq:full_vlasov_enskog_eq} can be also employed for the phase transition phenomena which include evaporation/condensation, among others \cite{grmela1971kinetic}. \\ \ \\
Similar to the Boltzmann collision operator,  $S^{\text{Ensk}}(\mathcal{F})$ describes the contribution of binary collisions to the evolution of MDF, however including dense effects. Therefore, the Enskog collision operator takes into  account the physical dimensions of the particles, which result in evaluating the distribution of the colliding pair, at different locations. Moreover, with respect to the Boltzmann operator here the collision rate is increased by the factor $Y$ which was originally justified considering the fact that less vacant physical space is available in a dense medium. The above-mentioned modifications of the Boltzmann operator lead to   
\begin{flalign}
S^{\text{Ensk}}(\mathcal{F})
= \frac{1}{m}& \int_{\mathbb{R}^3} \int_{0}^{2\pi} \int_{0}^{+\infty}  
 \Big [ Y (\bm{x}+\frac{1}{2} \sigma \hat{\bm{k}}) \mathcal{F}(\bm{v}^*, \bm{x})\mathcal{F}(\bm{v}^*_1,\bm{x}+\sigma \hat{\bm{k}})
\nonumber \\
 &-Y (\bm{x}-\frac{1}{2} \sigma \hat{\bm{k}})\mathcal{F}(\bm{v}, \bm{x})\mathcal{F}(\bm{v}_1, \bm{x}-\sigma \hat{\bm{k}}) \Big ] 
g { \color{referee2} \mathcal{H}(\bm g \cdot \hat{\bm{k}}) } \hat{b} d\hat{b} d\hat{\epsilon}  d^3 \bm{v}_1~;
\label{eq:enskog_op}
\end{flalign}
see  \cite{enskog1922kinetische,Chapman1953} for details. 
\\ \ \\
Note that the superscript $(.)^*$ indicates the post-collision state, $g=|\bm v_1 - \bm v|$ denotes the magnitude of the relative velocity of the colliding pair with velocities $(\bm v,\bm v_1)$ and $\hat{\bm k}$ is the unit vector connecting their centers in the physical space. Furthermore, $\hat{b}$ and $\hat{\epsilon}$ are the impact parameter and scattering angle, respectively {\color{referee2} and $\mathcal{H}(.)$ indicates the Heaviside step function}. A closure for the correlation function can be obtained e.g. by the Carnahan-Starling equation of state 
\begin{flalign}
Y &= \frac{1}{2}\frac{2-\eta}{(1-\eta)^3} \hspace{0.5cm} \text{and}\\
\eta &= nb/4,
\end{flalign}
where $b:=2 \pi \sigma^3/3$ is the second virial coefficient \cite{carnahan1969equation}. \\ \ \\
{\color{referee2}
Note that a more accurate description known as the \textit{Modified} Enskog equation, can be obtained by replacing $Y(\bm{x}\pm\frac{1}{2} \sigma \hat{\bm{k}})$ with an exact local-equilibrium pair-correlation function which includes higher order spatial dependencies of the density \cite{van1973modified,resibois1978h}. However, for simplicity, here we restrict the study to the conventional Enskog equation.
}
\subsection{Boundary Conditions}
\label{sec:bc}
\noindent In order to describe the evolution of the distribution $\mathcal{F}$, the Enskog-Vlasov equation needs to be equipped with appropriate boundary conditions. Note that here in comparison to the Boltzmann/Enskog equations, further information should be provided with regard to the behaviour of the molecular potentials at the boundaries. Therefore, besides introducing a boundary kernel for characterizing $\mathcal{F}$ conditional on the incoming particles, one should make assumptions on how the long-range interactions are affected by the boundaries of the physical domain. For the former problem, {\color{referee2} there exist} the standard Maxwell type boundary kernels for solid surfaces and flux boundary conditions for the open ones, with details that can be found e.g. in \cite{cercignani2000rarefied}. For the latter, we make a simplifying assumption that the physical boundary over which the Vlasov integral is computed, is unbounded. Therefore the Vlasov integral reads
\begin{eqnarray}
U(\bm x,t)&=&\int_{r>\sigma,\ \bm y \in {\mathbb{R}^3}}\phi(r)n(\bm y,t)d^3 \bm y; \ \ \ \ (\forall \bm x\in \mathbb{R}^3),
\end{eqnarray}
subject to the far-field condition $|\bm \xi (\bm x,t)|\to 0$ as $|\bm x|\to \infty$. Note that similar boundary conditions are employed for Coulomb interactions arising in plasmas \cite{birdsall2004plasma}. \\ \ \\
In order to achieve the far-field condition in a practical simulation, consider a given computational domain $\Omega$. Furthermore, let us decompose $\Omega^b=\mathbb{R}^3/ \Omega$  into the minimum number of disjoint sub-domains $\Omega^b_i$ i.e. $\Omega^b=\textrm{argmin}_{n}\cup_{i=1}^n\Omega^b_i$, where each $\Omega^b_i$ is a simply connected space and $\Omega^b_{i}\cap \Omega^b_{j}=\varnothing$ for $i\neq j$. Now by assigning the number density of each $\Omega^b_i$ to a uniform value $C_i(t)\in \mathbb{R}^+$, the desired condition $|\bm \xi (\bm x,t)|\to 0$ is obtained at $|\bm x|\to \infty$. Finally given the fact that $\phi(.)$ decays sharply in space, including just a few layers of ghost cells is practically sufficient to represent each $\Omega^b_i$.
\subsection{Stochastic Particle Methods}
\noindent In the framework of stochastic particle methods, the Enskog operator can be numerically approximated via Consistent Boltzmann Algorithm (CBA) \cite{Alexander1995}, Enskog Simulation Monte-Carlo (ESMC) \cite{Montanero1996,montanero1997simulation,montanero1997viscometric}, Frezzotti's algorithm \cite{Frezzotti1997} or the dense gas Fokker-Planck model (DFP) \cite{sadr2017continuous}. 
In the following section, numerical challenges associated with the Vlasov integral are investigated and a novel numerical scheme for an efficient computation of long-range interactions is introduced. Yet before proceeding, let us introduce the particle related variables necessary for our sequel discussion. Suppose $\bm M (t,\omega), \bm X(t,\omega) \in \mathbb{R}^3$ are random variables owing to the velocity and physical spaces, respectively. Here $\omega$ denotes a fundamental random event. Let $\bm M^{(i)}(t)$ and $\bm X^{(i)}(t)$ be realizations of $\bm M(t,\omega)$ and $\bm X(t,\omega)$, respectively. Therefore, in the stochastic particle framework, MDF reads
\begin{eqnarray}
\mathcal{F}(\bm v, \bm x, t)&=& \mathcal{M}\lim_{N_p\to \infty}\frac{1}{N_p}\sum_{i=1}^{N_p} \delta(\bm M^{(i)}(t)-\bm v)\delta(\bm X^{(i)}(t)-\bm x),
\label{eq:dist-particle}
\end{eqnarray}
where $\mathcal{M}$ is the total mass in the system and $\delta(.)$ is the Dirac delta measure. In practice, a finite number of particles $N_p$ can be employed and hence the stochastic representation becomes subject to the statistical errors, including both bias and noise.
\section{Treatment of the Vlasov integral}
\label{sec:treatment}
\noindent In comparison to the Enskog equation, here an extra challenge of approximating the Vlasov integral arising from long-range interactions has to be dealt with. 
Besides direct approximations of the Vlasov integral with a quadrature rule or a particle Monte-Carlo method, two alternatives are considered. First, a density expansion method introduced in \cite{he2002thermodynamic} is reviewed. The method makes use of the fact that once the density variation in the physical domain is not too large, an approximation of the Vlasov integral becomes possible by adopting an expansion technique. In particular, through convolution of the molecular potential $\phi$ with respect to the Taylor expanded number density $n$, the Vlasov integral can be represented as a functional of the density and its derivatives. 
Next, we provide a novel methodology, where the Vlasov integral is translated to its corresponding partial differential equation (PDE) of a screened-Poisson type. Motivated by treatment of the Coulomb interactions in the Vlasov-Fokker-Planck description of plasmas \cite{thomas2012}, here the solution of the Vlasov integral can be computed by a screened-Poisson solver.
 In the following, each of the above-mentioned three approaches are discussed, theoretically justified and the corresponding numerical schemes are provided. 
\subsection{Direct method}
\label{sec:direct_method}
\noindent In the framework of particle methods by using the identity \eqref{eq:dist-particle}, the Vlasov integral simplifies to
\begin{eqnarray}
U(\bm x,t)&=&\frac{1}{m}\int_{\mathbb{R}^3} \int_{r>\sigma}\phi(r)\mathcal{F}(\bm v, \bm y,t) d^3 \bm y d^3 \bm v \nonumber \\
&=&\lim_{N_p\to \infty}{\frac{\mathcal{M}}{N_p m} }\sum_{i=1}^{N_p}\int_{\mathbb{R}^3}\int_{r>\sigma} \phi(r)\delta(\bm y-\bm X^{(i)}(t)) \delta (\bm v-\bm M^{(i)}(t)) d^3 \bm y d^3 \bm v \nonumber \\
&=& \omega_s \lim_{N_p\to \infty} \sum_{i=1}^{N_p}\int_{r>\sigma} \phi(r)\delta(\bm y-\bm X^{(i)}(t))  d^3 \bm y \ \ \ \ \ \left(\omega_s:=\frac{\mathcal{M}}{N_pm} \right) \nonumber \\
&=&\omega_s\lim_{N_p\to \infty} \sum_{i=1}^{N_p}\phi\left(|\bm x-\bm X^{(i)}(t)|\right) \ \ \ \ \textrm{s.t.} \ \ \ \ |\bm x-\bm X^{(i)}(t)|>\sigma\, \label{eq:U-part}
\end{eqnarray}
where $\omega_s$ is the statistical weight of each particle.
However note that the attractive part of the potential $\phi( r)$ decays very sharply with respect to $|\bm x-\bm X^{(i)}|$, and hence the sum is controlled mainly by the particles in the vicinity of $\bm x$. Thus, in order to ensure a sufficient statistical resolution around each point in the physical domain, a huge number of particles have to be employed. It is important to note that while Eq.~\eqref{eq:U-part} results in a $\mathcal{O}(N_p^2)$ computational cost in a brute-force approach, the cost can be significantly reduced by using fast multipole ideas (see e.g. \cite{rokhlin1985rapid,engheta1992fast,coifman1993fast}).\\ \ \\
\noindent A less computationally intensive approximation can be obtained by calculating the Vlasov integral with a quadrature rule. Suppose $\left(\bm y^{(i)},i\in \{1,...,N_c\}\right)$ is collection of points for which the integrand has to be computed, therefore we get  
\begin{eqnarray}
U(\bm x,t) \approx  \sum_{i=1}^{N_c} \phi\left(|\bm x- \bm y^{(i)}|\right)n(\bm y^{(i)},t)w^{(i)} \ \ \ \ \textrm{s.t.} \ \ \ \ |{\color{referee2} \bm x}-\bm y^{(i)}|>\sigma,
\label{eq:quad}
\end{eqnarray}
where $w^{(i)}$ corresponds to the quadrature weights. Since in practice variations of $\phi(.)$ are much sharper than $n(.,t)$, advantage can be taken by using a coarse scale approximation of the latter in Eq.~\eqref{eq:quad}.  As a result, for a piecewise constant coarse scale approximation, the quadrature approach can lead to cheaper computations in comparison to Eq.~\eqref{eq:U-part}. \\ \ \\
For treatment of open boundaries outlined in \S \ref{sec:bc}, each domain $\Omega_i^b$ is represented by a number of ghost cells according to a cut-off value $r_{cut}$. The number density of each group of ghost cells then are assigned to the spatially averaged number density value computed from the neighbouring cells inside the domain.

\subsection{Density expansion}
\label{sec:density_expansion}
\noindent For a number density field with slight variations, one can obtain a simple alternative to the Vlasov integral \cite{he2002thermodynamic}. Note that the integral can be seen as convolution of $\phi$ and $n$
\begin{flalign}
U(\bm x, t) &= \int_{r>\sigma} \phi(|\bm y - \bm x|) n(\bm y, t) d^3 \bm y\nonumber \\
&=   \int_{|\bm y|>\sigma} \phi(|\bm y| ) n(\bm y-\bm x, t) d^3 \bm y~.
\end{flalign}
Suppose the number density is a smooth field, and thus using the Taylor expansion around $\bm x$ leads to
\begin{flalign}
U(\bm x, t) \approx & n(\bm x,t) \underbrace{ \int_{|\bm y|>\sigma} \phi(|\bm y| )d^3 \bm y}_{I_0}
-   \frac{\partial n (\bm x,t)}{\partial x_i} \underbrace{\int_{|\bm y|>\sigma} \phi(|\bm y| ) y_i d^3 \bm y}_{I_1} \nonumber \\
&+ \frac{1}{2}\frac{\partial^2 n (\bm x,t)}{\partial x_i \partial x_j}  \underbrace{ \int_{|\bm y|>\sigma} \phi(|\bm y|) y_i y_j d^3 \bm y}_{I_2} ~, \label{eq:den-exp}
\end{flalign}
where derivatives of $n(.,t)$ higher than second-order are neglected.
The integral term with the first-order derivative of $n$ vanishes since $I_1$ is over an odd function in $\mathbb{R}^3 \setminus  \{ {B}(\bm 0, \sigma) \}$ for the ball $B(\bm 0, \sigma)$ centred at the origin, of the radius $\sigma$. Similarly, the off-diagonal  terms in the integral $I_2$ become zero which together with the analytic solution of $I_0$  yield
\begin{flalign}
U \approx & 
4 \pi \phi_0 \sigma^3\Bigg( \frac{n}{3}+\frac{\sigma^2}{2} \frac{\partial^2 n }{\partial x_j \partial x_j} \Bigg) ~.
\end{flalign}

\noindent Even though the method of density expansion reviewed here is very attractive due to its simplicity, it requires somewhat strong assumptions underlying the number density $n$. Besides the requirement of $n(.,t)\in C^\infty$ for the Taylor expansion, note that $I_p \to \infty$ for $p\ge 6$.
Therefore unless all derivatives of $n(.,t)$ with the order higher than $5$ vanish, the truncated expansion \eqref{eq:den-exp} is not well defined. Nevertheless,  for practical computations, one can attain a smooth estimate of the number density along with its derivatives e.g. via the method of Smoothed-Particle-Hydrodynamics (SPH) \cite{monaghan1992smoothed} as briefly mentioned below. 
 For a more generic kernel based estimation of derivatives see \cite{kuchlin}. \\ \ \\
 \noindent For simplicity, let us employ the Gaussian kernel 
\begin{eqnarray}
W( r, h) &:= &\frac{1}{h^3 (2 \pi)^{3/2}} e^{-r^2/(2h^2)},
\end{eqnarray}
 where $h$ denotes the smoothing length. 
Following the SPH methodology, the number density at each point $\bm x$  can be estimated via 
\begin{flalign}
n(\bm x,t) \approx \omega_s \sum_{i=1}^{N_p}  W(|\bm x-\bm X^{(i)}(t)|, h)
\end{flalign}
 such that as $N_p \rightarrow \infty$ and $h\to 0$, the equality  is obtained.  As a result, one can compute derivatives of the smoothed number density, e.g.
\begin{flalign}
 \frac{\partial^3 n (\bm x,t)}{\partial x_i \partial x_j \partial x_j } 
 &\approx
  \omega_s \sum_{k=1}^{N_p}   W(|\bm x-\bm X^{(k)}(t)|, h)   \frac{A_{i}^k}{h^4}
  \end{flalign}
  with
  \begin{flalign}
  A_{i}^k &:= 3(x_i-X^{(k)}_i(t)) -\frac{1}{h^2} (x_i-X^{(k)}_i(t)) (x_j - X^{(k)}_j(t)) (x_j - X^{(k)}_j(t))~.
\end{flalign}
Similar to the direct method, the boundary conditions for the long-range forces are achieved by introducing the ghost cells for each $\Omega_i^b$. Here the cut-off radius (and hence number of ghost cells) is found from support of the kernel, which is used for estimating derivatives of the number density.
\subsection{Screened-Poisson equation}
\label{sec:screened_poisson_solution}
\noindent In order to avoid stiff resolution requirements in the direct method mentioned in \S~\ref{sec:direct_method} or otherwise strong assumptions on the number density required by the method described in \S~\ref{sec:density_expansion}, here we introduce an efficient alternative by transforming the Vlasov integral to a screened-Poisson PDE. The transformation is carried out  using the method of fundamental solutions. \\


\noindent The basic idea is approximating the expression for attractive part of the Sutherland potential $\phi(\bm r)$ with a function which is the fundamental solution of the screened-Poisson equation. \\ \ \\
Let us approximate
  $\phi( r)$ by an estimate $\tilde{\phi}( r)$ defined as
\begin{flalign}
\tilde{\phi}( r) &= a G( r) \hspace{0.5cm} \text{and}\\
G( r) &= \dfrac{e^{-\lambda r}}{4 \pi r}.
\end{flalign}
{\color{referee1} The coefficients $a$ and $\lambda$ are to be fixed by minimizing an appropriate error norm. Since only the gradient of the potential appears in the evolution equation, we consider the minimization problem $ || \partial_r \phi(r)-\partial_r \tilde{\phi}(r)   ||_2   $ for $r \in (\sigma, \infty)$. Using the method of  non-linear least squares or gradient descent, the optimal values for $a$ and $\lambda$ can be obtained}.
 It is interesting to see that by setting $\lambda=0$, the Coulomb law is recovered. \\ \ \\
{\color{referee1} 
One of the important macroscopic laws in thermo-fluid phenomena is the equation of state. Hence it can serve as a relevant measure to characterize the error committed by our approximate potential. Consider a decomposition of the equilibrium pressure to attractive and repulsive parts \cite{karkheck1981kinetic,Hirschfelder1963}
\begin{flalign}
p_{\text{eq}} (n,T) &= nk_bT(1+nbY) + p^{\text{att}}_{\text{eq}}(n)
\hspace{0.5cm} \text{and}\\
p_{\text{eq}}^{\text{att}}  (n)
&= -\frac{2 \pi}{3} n^2  \int_{\sigma}^{\infty}     \frac{\partial \phi(r)}{\partial r} r^3 dr~,
\end{flalign} 
where $k_b$ indicates the Boltzmann constant and $T$ is the kinetic temperature.
The attractive part of the equilibrium pressure becomes
\begin{flalign}
\label{eq:stateSuth}
p^{\text{att, Suth}}_{\text{eq}} (n) &=   - \frac{4 \pi}{3}  \phi_0  \sigma^3 n^2
\end{flalign}
for the Sutherland potential, and
\begin{flalign}
\label{eq:stateSC}
p^{\text{att, SC}}_{\text{eq}} (n) &=   \frac{a }{6} \frac{ e^{-\lambda \sigma} }{ \lambda^2 }(\lambda^2 \sigma^2 + 3 \sigma \lambda +3 ) n^2~
\end{flalign}
for the screened-Coulomb approximation. Equations \eqref{eq:stateSuth} and \eqref{eq:stateSC} give us an explicit comparison between the two potentials. For example in the case of argon, the resulting approximation of $\phi(r)$ by $\tilde{\phi}(r)$  provides us with a relative error of $\mathcal{O}(10^{-8})$ in $p_{eq}^{att}$. 
} 
 \\ \ \\
Consider the screened-Poisson (modified Helmholtz) equation
\begin{eqnarray}
\label{eq:screen-unb}
\left(\Delta-\lambda^2\right)u(\bm x,t)&=&n(\bm x,t); \ \ \ \ (\forall \bm x\in \mathbb{R}^3)
\end{eqnarray}
subject to the condition $|\partial_{x_i} u(\bm x,t)|\to 0 $ as $|\bm x|\to \infty$, where $\Delta$ is the Laplacian. It is a classic result in theory of fundamental solutions for Helmholtz type equations that
the solution of Eq.~\eqref{eq:screen-unb} becomes
\begin{eqnarray}
u(\bm x,t)&=&\int_{\mathbb{R}^3}G(r)n(\bm y,t)d^3 \bm y,
\end{eqnarray}
see e.g. \cite{kythe2012fundamental,caprino2002two}. 
Therefore, for unbounded domains discussed in \S.\ref{sec:bc}, the solution of the screened-Poisson equation is related to  the solution of the Vlasov integral  through
\begin{flalign}
\label{eq:integ-sp}
U(\bm x, t)   &=  \int_{r>\sigma}\phi(  r) n(\bm y, t) d^3 \bm y \nonumber \\
&\approx  a\int_{r>0}G( r) n(\bm y, t) d^3 \bm y
   -   a\underbrace{\int_{r<\sigma}G( r) n(\bm y, t) d^3 \bm y}_{\tilde{U}_{r<\sigma}}\nonumber \\
   &= a\ u(\bm x) - a\ \tilde{U}_{r<\sigma}~.
\end{flalign}
Assuming that the number density remains constant for $r \in (0, \sigma]$, we get
\begin{flalign}
\tilde{U}_{r < \sigma} (\bm x)
&\approx
n(\bm x) \int_{0}^{\sigma}\int_{0}^{2\pi} \int_{0}^\pi G( r) r^2 \sin \theta\  d\theta d\phi dr \nonumber \\
&=
n(\bm x) \Big( \frac{1}{\lambda^2} - \frac{(\sigma \lambda+1)e^{-\sigma \lambda}}{\lambda^2} \Big).
\end{flalign}
Despite the fact that the considered Vlasov integral is defined on $\mathbb{R}^3$, in practice one needs a bounded domain equipped with appropriate boundary conditions in order to solve the screened-Poisson equation. Before proceeding, let us consider $\Omega \subset \mathbb{R}^3$ with the boundary $\partial \Omega$, for which
\begin{eqnarray}
\left(\Delta-\lambda^2\right)\psi(\bm x,t)&=&n(\bm x,t) \ \ \ \ \ (\forall \bm x\in\Omega) \ \ \ \textrm{and} \label{eq:sp-bc}\\
\psi(\bm y,t)&=&g(\bm y,t) \ \ \ \ \ \ (\forall \bm y\in \partial \Omega). \label{eq:dir-bc}
\end{eqnarray}
To better understand  the relation between $\psi(\bm x,t)$ and $u(\bm x,t)$, note that the latter is the solution of the screened-Poisson equation in $\mathbb{R}^3$, while the former is for $\Omega$. It is easy to see that due to the solution uniqueness  of Eq.\eqref{eq:sp-bc} subject to the boundary condition given by Eq.~\eqref{eq:dir-bc}, we have
\begin{eqnarray}
u(\bm x,t)&=& \psi(\bm x,t) \ \ \ \ (\forall \bm x\in \Omega)
\end{eqnarray}
provided
\begin{eqnarray}
\label{eq:bc-consist}
g(\bm y,t)&=&u(\bm y,t) \ \ \ \ (\forall \bm y\in \partial \Omega).
\end{eqnarray}
Therefore once $u$ is computed on the boundary $\partial \Omega$, Eq.~\eqref{eq:bc-consist} ensures the consistency between the solution of the screened-Poisson in $\Omega$ and the one corresponding to $\mathbb{R}^3$, for every $\bm x\in \Omega$. \\ \ \\
Hence, the procedure for computing the Vlasov integral based on the screened-Poisson equation includes two steps:
\begin{enumerate}
\item {\it Calculating the boundary condition.} Here one can rely on the direct method described in \S \ref{sec:direct_method}, in order to first compute $U(\bm x,t)$ for $\bm x\in \partial \Omega$. Notice that this step is relatively cheap since we compute the integral only for elements on the boundary. Next, the field potential $u(\bm x,t)$ is found through Eq.~\eqref{eq:integ-sp}, closing the boundary condition Eq.~\eqref{eq:bc-consist} required for solving Eq.~\eqref{eq:sp-bc}. 
\item {\it Discretizing the screened-Poisson equation.} This step considers discretization of Eq.~\eqref{eq:sp-bc}, subject to the above-mentioned Dirichlet boundary condition. Let the subscript $(\ .\ )^t_{i_1,i_2,i_3}$ denote the approximated value computed based on a given numerical scheme on a node with the index $(i_1,i_2,i_3)$ at time $t$.  A simple finite difference discretization then leads to 
\begin{flalign}
\frac{\psi^t_{i+1,j,k}+\psi^t_{i-1,j,k}}{\Delta x^2} 
&+ \frac{\psi^t_{i,j+1,k}+\psi^t_{i,j-1,k}}{\Delta y^2} 
+ \frac{\psi^t_{i,j,k+1}+\psi^t_{i,j,k-1}}{\Delta z^2}  \nonumber \\
&-\psi^t_{i,j,k}\left( \frac{2}{\Delta x^2} + \frac{2}{\Delta y^2} + \frac{2}{\Delta z^2} + \lambda^2 \right) 
= -n^t_{i,j,k}~,
\end{flalign}
where $\Delta x$, $\Delta y$ and $\Delta z$ indicate distances between adjacent nodes along $x_1$, $x_2$ and $x_3$ directions, respectively.\\ \ \\
It is important to emphasize that the screening coefficient $\lambda$ improves the dominance of the main diagonal associate with the coefficient matrix. Similar to the classical Poisson problem,  the linear system of equations corresponding to the  discrete formulation of the screened-Poisson equation can  then be solved e.g. by direct  or iterative solvers including Multi-Grid and  Krylov space methods \cite{saad2003iterative}.
\end{enumerate}
\noindent The main advantage of the screened-Poisson approach lies on the fact that the N-body problem implied by the Vlasov integral is reduced to an elliptic PDE. As a result, we avoid the stringent resolution in short distances i.e. $r \in (\sigma, 3 \sigma)$ required by the direct computation of the Vlasov integral. Furthermore, there is no need to introduce a cut-off for the potential kernel $\phi$. 

\section{Results}
\label{sec:results}
\begin{figure}
  \centering
  \scalebox{1}{\includegraphics[ angle =0]{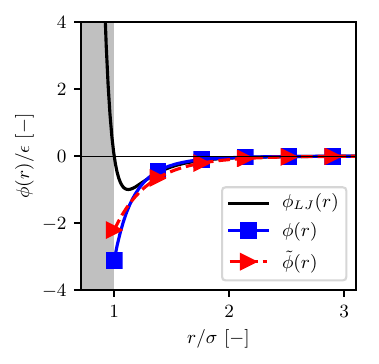}}
  \caption{ {\color{referee1} The Lennard-Jones molecular potential compared to the fitted Sutherland and the screened-Coulomb potentials.}}
  \label{fig:potentials}
\end{figure}

\noindent In this section, performances of all three approaches i.e. the direct method (\S~\ref{sec:direct_method}), density expansion (\S~\ref{sec:density_expansion}) and screened-Poisson formulation (\S~\ref{sec:screened_poisson_solution}) are compared with respect to each-other, in terms of both accuracy and computational efficiency. Two test cases are considered. First, the well-known planar Couette flow is investigated at the dense transitional regime. Next, evaporation of a liquid slab exposed to the vacuum is studied, below its critical point. Note that for both cases the short-range interactions are described by the Enskog collision operator, where the ESMC method is employed. The benchmark is provided by the direct calculation of the Vlasov integral described in \S~\ref{sec:direct_method}.  Furthermore, results of MD simulations using Leapfrog integration are presented for the evaporation scenario \cite{rapaport2004art} . \\ \ \\
 {\color{referee1_2nd} Using argon, the Sutherland coefficient  $\phi_0 = 3.115249 \epsilon $  with the effective diameter $\sigma$ were chosen based on a curve fit to the attractive part of the Lennard-Jones potential \cite{rowley1975monte}
\begin{flalign}
\phi_{\text{LJ}}(r) = 4 \epsilon \left[  \left(\frac{\sigma}{r}\right)^{12} - \left(\frac{\sigma}{r}\right)^6 \right]~,
\end{flalign}
for $r \in (\sigma_{\text{min}},\infty)$ with $\epsilon =  119.8 k_b$, $\sigma = 3.405\times 10^{-10} \ \mathrm{m}$ \cite{rowley1975monte}  and $\sigma_{\text{min}} = 2^{1/6} \sigma$.} Moreover, the resulting minimization problem for the screened-Coulomb potential leads to {\color{referee1_2nd} $a=-1.64835851\times 10^{-28} \ \mathrm{N.m}^2$ and $\lambda=6.91304716\times 10^{9} \ \mathrm{m}^{-1}$}. The employed potentials are depicted in Fig.~\ref{fig:potentials}.
 The molecular mass is set to $m=6.6335214\times 10^{-26} \ \mathrm{kg}$. Note that {\color{referee1} for kinetic based simulations,} the hard sphere model with the diameter $\sigma$ is adopted for short range collisions. 
 \\
{\color{referee1}
The macroscopic quantities are obtained by sampling the velocity moments. Consider a function $\bm Q(.)$, and let
\begin{flalign}
\left\langle \bm{{Q}} ( \bm v ) \right\rangle := 
\int_{\mathbb{R}^3} \bm{Q}( \bm v  )  \mathcal{F} (\bm v , \bm x, t ) d^3 \bm v ~.
\end{flalign}
Through the identity \eqref{eq:dist-particle}, the expectation can be estimated
\begin{flalign}
\langle \bm Q(\bm v) \rangle
\approx
\frac{1}{\delta \Omega^{(j)}} \sum_i w \bm Q ( \bm{M}^{(i)} ),\ \ \ \ \left(\bm{X}^{(i)},\bm x\right) \in \Omega^{(j)}~;
\end{flalign}
\noindent where $\Omega^{(j)}$ indicates a computational cell in the physical space, $\delta \Omega^{(j)}$ denotes its volume and $w:=\mathcal{M}/N_p$ is the statistical weight of each particle indexed by $i\in\{1,...,N_p\}$. Note that for simplicity an equidistant Cartesian grid was employed in this study.}

%
%
%
%
%
%
%
%
%

\subsection{Couette Flow}
\begin{figure}
  \centering
  \begin{subfigure}[b]{0.48\columnwidth}
  \scalebox{1}{\includegraphics[ angle =0]{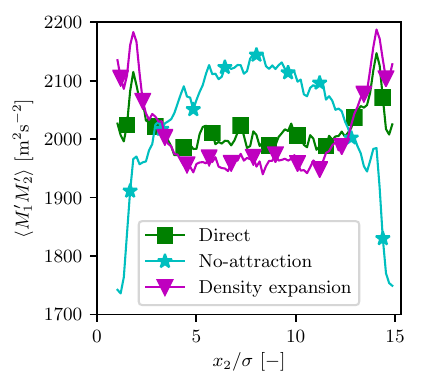}}
  \end{subfigure}
  \  \  \ 
  \begin{subfigure}[b]{0.48\columnwidth}
   \scalebox{1}{\includegraphics[ angle =0]{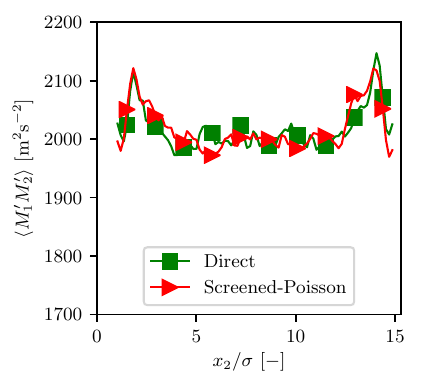}}
  \end{subfigure}
  \
    \begin{subfigure}[b]{0.48\columnwidth}
  \scalebox{1}{\includegraphics[ angle =0]{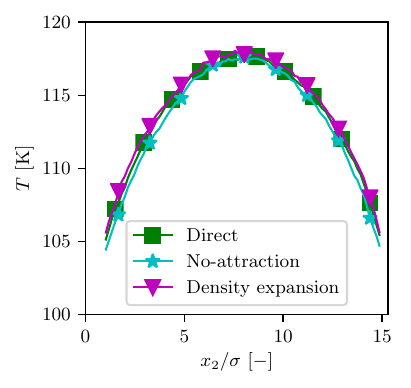}}
  \end{subfigure}
  \  \  \ 
  \begin{subfigure}[b]{0.48\columnwidth}
   \scalebox{1}{\includegraphics[ angle =0]{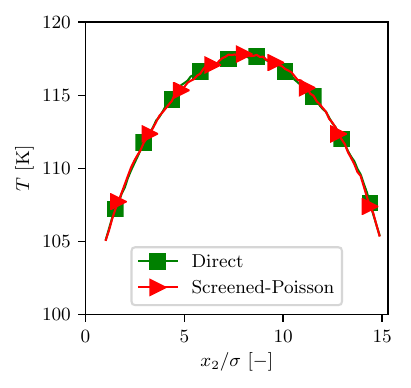}}
  \end{subfigure}
  \caption{Temperature and shear profiles, for the Couette flow at $\mathrm{Kn} = 0.1$ and $n_0b=0.7$. The direct method, density expansion and screened-Poisson formulation were employed together with  ESMC in order to compute both collision as well as the attractive part of the potential. For comparison, pure ESMC computations labeled with no-attraction, are depicted as well.}
  \label{fig:couette_M1M2_T}
\end{figure}

\noindent The influence of attractive forces in non-equilibrium regime is investigated by simulating a planar Couette flow. Initially, the fluid is at rest with the temperature $T_0=100$~K and the number density $n_0$ such that $n_0b = 0.7$. The physical boundaries are composed of two parallel walls, distanced by $L$ resulting in $\mathrm{Kn}=\lambda_0/L=0.1$, where $\lambda_0$ is the mean free path computed at $n_0$. The moving walls with the velocities $U_w = (\pm 150, 0, 0)^T \mathrm{ms^{-1}}$, yield non-equilibrium flow quantities varying along the $x_2$-direction, normal to the walls.
{\color{referee1}  
Fully diffusive isothermal walls at the temperature $T_w=100 \ \mathrm{K}$ are employed \cite{Bird}, where the particle-wall collision event is considered once the shortest distance between the boundary and the center of the particle becomes smaller than the effective diameter.} \\ \ \\
A refined grid of $100$ cells, each of which initially populated by $1000$ particles is employed. The time  step size $\Delta t =   1.938\times 10^{-14} \ \mathrm{s}$ is chosen as a small fraction of the mean collision time. A statistically stationary condition was obtained after around $10^4\  \Delta t$. Afterwards, time averaging was used in order to control the statistical error. \\ \ \\
A cut-off value of $3\sigma$ was applied both inside the physical domain and at the boundaries, resulting in roughly $20$ ghost cells for each boundary. Note that the ghost cells inside each wall boundary possess a uniform number density equals to the adjacent cell (resulting in the Neumann boundary condition for the number density). 
\begin{figure}
  \centering
  \begin{subfigure}[b]{0.48\columnwidth}
  \scalebox{1}{\includegraphics[ angle =0]{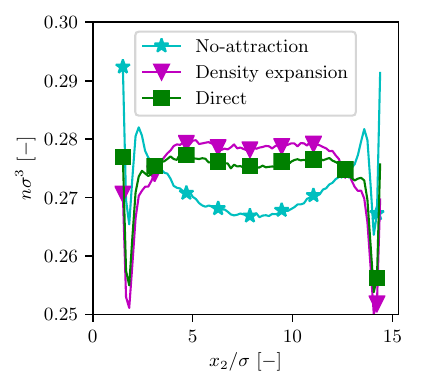}}
  \end{subfigure}
  \  \  \ 
  \begin{subfigure}[b]{0.48\columnwidth}
   \scalebox{1}{\includegraphics[ angle =0]{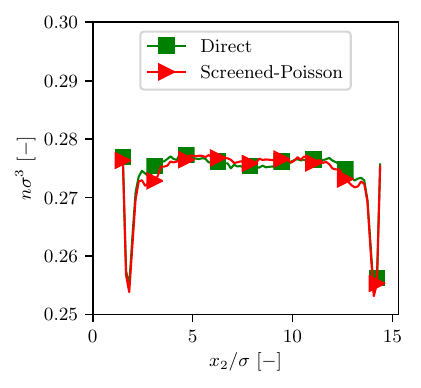}}
  \end{subfigure}
  \
    \begin{subfigure}[b]{0.48\columnwidth}
  \scalebox{1}{\includegraphics[ angle =0]{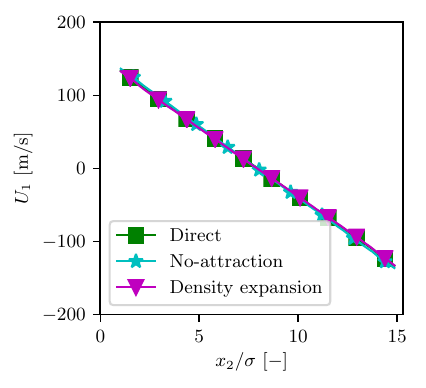}}
  \end{subfigure}
  \  \  \ 
  \begin{subfigure}[b]{0.48\columnwidth}
   \scalebox{1}{\includegraphics[ angle =0]{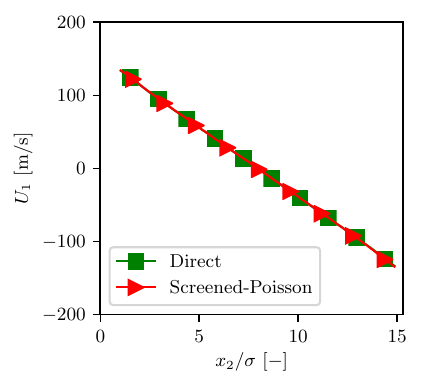}}
  \end{subfigure}
  \caption{Normalized number density $n \sigma^3$ {\color{referee1} and bulk velocity $U_1$}, for the Couette flow at $\mathrm{Kn} = 0.1$ and $n_0b=0.7$. The direct method, density expansion and screened-Poisson formulation were employed together with  ESMC in order to compute both collision as well as the attractive part of the potential. For comparison, pure ESMC computations labeled with no-attraction, are depicted as well.}
  \label{fig:couette_nsigma3}
\end{figure}
\\ \ \\
\noindent The normalized number density is shown for ESMC simulations in Fig.~\ref{fig:couette_nsigma3}, where the long-range interactions were computed using the direct method, density expansion and screened-Poisson. It is interesting to see that the density variation without attractive part of the potential (i.e. pure ESMC) is more pronounced in comparison to the full Enskog-Vlasov kinetics. Moreover, the temperature $T$ and the kinetic shear stress $\langle M^\prime_1  M^\prime_2 \rangle$ are shown in Fig.~\ref{fig:couette_M1M2_T}. Note that $\bm M^\prime$ indicates {\color{referee2} the} fluctuation velocity and $\langle...\rangle$ represents the expectation. While overall both the density expansion and the screened-Poisson formulation lead to predictions close to the direct method, slight yet systematic errors in the density expansion results can be observed.


\subsection{Evaporation}
\begin{figure}
  \centering
  \begin{subfigure}[b]{0.48\columnwidth}
  \scalebox{1}{\includegraphics[ angle =0]{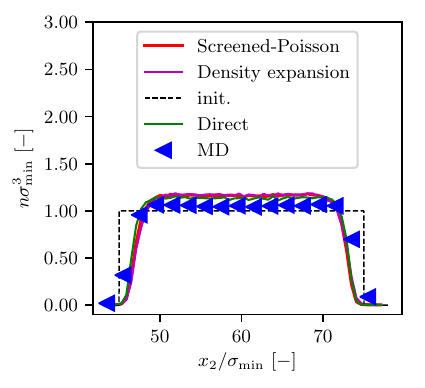}}
  \end{subfigure}
  \  \  \ 
  \begin{subfigure}[b]{0.48\columnwidth}
   \scalebox{1}{\includegraphics[ angle =0]{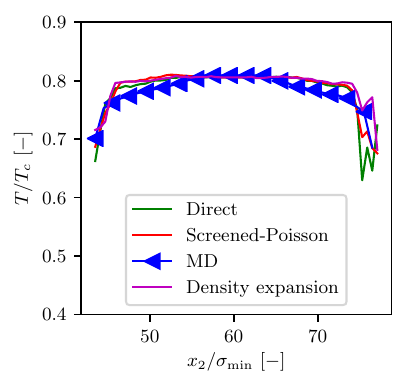}}
  \end{subfigure}
  \
  \begin{subfigure}[b]{0.48\columnwidth}
  \scalebox{1}{\includegraphics[ angle =0]{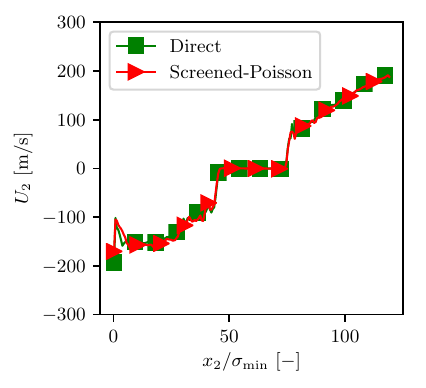}}
  \end{subfigure}
  \  \  \ 
  \begin{subfigure}[b]{0.48\columnwidth}
   \scalebox{1}{\includegraphics[ angle =0]{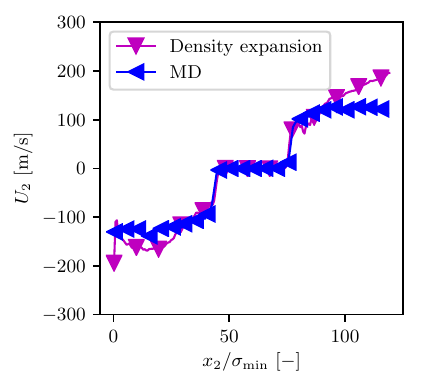}}
  \end{subfigure}
  \caption{Normalized number density{\color{referee1}, bulk velocity and temperature profiles}  for $x_2 \in [\frac{5L}{4}, \frac{11L}{4}]$ averaged over $t \in [t_0,t_f]$, computed for liquid argon evaporation to vacuum at $T_{\mathrm{init}} = {\color{mine}0.8} \ T_C$. The direct method, density expansion and screened-Poisson formulation were employed together with  ESMC in order to compute both collision as well as the attractive part of the potential. Furthermore, MD results are provided.
  }
  \label{fig:vacuum_n}
\end{figure}

\noindent For further validation of the introduced screened-Poisson formulation of long-range interactions, evaporation of the liquid argon at $T_L= {\color{mine}0.8 T_C}$ to the vacuum is investigated, where the critical temperature {\color{mine} $T_C = 124.1367 \  \mathrm{K}$} is considered. The setting is similar to the vacuum test case presented in \cite{frezzotti2005mean}. For a similar setting, an MD simulation is also conducted. 
{\color{referee1_2nd}
Similar to the mentioned study, here a thermostat of the length $4\sigma$ was deployed in the center of the liquid by scaling the particles velocities to maintain a constant temperature of $T_L$.
}
\begin{itemize}
\item[1-] \textit{ESMC-based simulations:} Consider a domain size of $1\times 4L \times 1 \  \mathrm{m}^3$, where the particles are initialized {\color{referee2} uniformly} inside the slab of the size $1\times L \times 1 \ \mathrm{m}^3$ with  $L =  30 {\color{referee1_2nd} \sigma_{\mathrm{min}}}$ and the initial number density $n_{0} = {\color{mine} 1.791157\times 10^{28}} \ \mathrm{m^{-3}}$. Also, the velocity of particles are initialized from the Maxwellian with the temperature $T_L$.  While the directions $x_1$ and $x_3$ are considered to be periodic, the evaporation happens along $x_2$ direction. Once a particle leaves the domain along the $x_2$-direction, it will be re-initialized inside the slab by a velocity according to the initial Maxwellian and a random position uniformly distributed inside the slab. Based on a convergence study, the physical domain is discretized by $160$ cells in $x_2$-direction. Initially $N_p = 2 \times 10^5$ particles were generated and the time step $\Delta t_{\mathrm{MC}}= {\color{mine} 4.770503} \times 10^{-14}$ s was adopted. According to the cut-off value $3\sigma$, one ghost cell at each boundary side of $x_2$ axis was employed. Note that since a relatively large boundary has been chosen, the results are not sensitive to the number density of the ghost cells. Here the average value of adjacent cells were fed into the ghost cells resulting in the number density of ${\color{mine} 2.01\times 10^{25}}$~${m^{-3}}$ for the ghost cells, at the steady state.  {\color{referee1} For the direct method , a refined spherical mesh with $N_{\text{fine}}=250,000$ discrete points was utilized for each computational cell. Note that similar discretisation was employed in computation of the Vlasov integral at the boundaries, arising in the screened-Poisson method. Once the Vlasov integral is calculated, its gradient is approximated by the finite volume method.}
\item[2-] \textit{MD simulation:}
In order to make the three dimensional MD simulation computationally feasible, smaller domain lengths in $x_1$ and $x_3$ are chosen, i.e. {\color{referee1}$0.5L\times 4L \times 0.5L \ \mathrm{m}^3$}, such that $N_p = {\color{referee1} 6750}$.
Initial and boundary conditions of the molecular velocity were chosen similar to the ESMC simulation. However, particles are initially placed in the slab with the equal distance $\sigma$ from each-other. To re-initialize a particle which leaves the domain, the center of a cell which is randomly chosen from empty volumes inside the slab, is used as the new position of the particle.  
Note that directions $x_1$ and $x_3$ are treated as periodic boundaries, while the $x_2$-direction is assumed to be Neumann at the boundaries. For stability reasons, a small time step of $\Delta t_{\mathrm{MD}} = {\color{mine} 1.125002 \times 10^{-14}}$ s has been deployed for the MD simulations.	
{\color{referee1}
To further control the statistical errors,  the ensemble averages were taken over 20 independent MD simulations.
}
\end{itemize}  
\noindent  
\noindent  After $ t_0 = {\color{mine} 0.1908}$ ns, the stationary distribution is achieved and time averaging was performed for $t_0 \leq t \leq t_f$, where $t_f =  {\color{mine} 0.38164}$ ns. Profiles of number density are illustrated in Fig.~\ref{fig:vacuum_n}.
The solution obtained based on the screened-Poisson problem shows a good agreement with the direct method.  Notice that the density expansion shows a slight over prediction for the number density profile. {\color{referee1_2nd} While overall good agreement between kinetic based and MD simulations are observed, the discrepancies are not negligible. This can be partly justified based on the difference in transport properties resulting from the Sutherland potential in comparison to the Lennard-Jones potential. More important, the thermostat implementation can also affect the results.}


\subsection{Computational complexity}
\label{sec:computational_complexity}
\begin{figure}
  \centering
  \begin{subfigure}[b]{0.48\columnwidth}
  \scalebox{1}{\includegraphics[ angle =0]{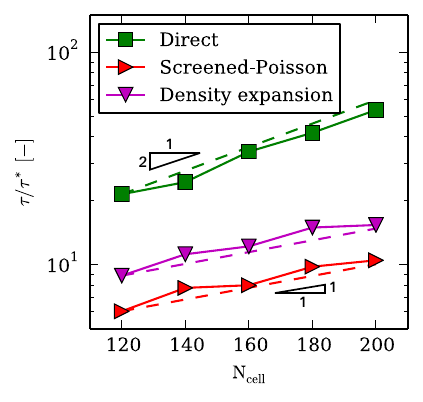}}
  \end{subfigure}
  \caption{Computational cost of the  evaporation test case using the direct method, density expansion and screened-Poisson formulation. Here $\tau/\tau^*$ denotes a normalized CPU time.}
  \label{fig:vacuum_complexity}
\end{figure}

\begin{enumerate}
\item In order to solve the Vlasov integral directly, one needs to discretize the physical space resolving $\sigma$. Furthermore in the brute-force approach, at each discrete point, the Vlasov integral should be estimated resulting in $\mathcal{O}(N^2_{\textrm{cell}})$ {\color{referee2} operations} where $N_\mathrm{cell}$ denotes  the number of computational cells. Certainly one can significantly accelerate the computations by using e.g. fast multipole approaches \cite{rokhlin1985rapid,engheta1992fast,coifman1993fast}. Moreover, since the number density varies on a relatively coarser scale, an alternative approach would be to sample the number density on a coarse mesh, whereas a finer spherical mesh with $N_\mathrm{fine}$ is adopted for $\phi$ on top of each coarse cell (see \S~\ref{sec:direct_method}). 
\item  In case of the density expansion approach, if a kernel based estimate is employed to compute the number density derivatives, the cost becomes proportional to the total number of particles $N_p$. In particular, depending on the support of the kernel, the complexity scales with $\mathcal{O}( N_{\textrm{p}}N_\mathrm{support})$, where $N_\mathrm{support}$ indicates the number cells within the support of the kernel. Moreover assuming that $N_p$ is proportional to $N_{\textrm{cell}}$ we get a linear cost proportionality with respect to the number of cells.
\item Analyzing the complexity associated with the screened-Poisson PDE depends very much on the applied method. Efficient Poisson type solvers gives rise to the costs of order $\mathcal{O}(N_\mathrm{cell}\log (N_{\mathrm{cell}}))$ or $\mathcal{O}(N_{\mathrm{cell}})$, depending on the number of dimensions \cite{saad2003iterative}. It is important to note that the number of cells required by the screened-Poisson solver is significantly smaller than the one required by the direct method, since here the computational cells have to only resolve variations of the number density (compared to the direct method which should resolve $\sigma$). 
\end{enumerate}
\noindent \sloppy For a fixed $N_{\textrm{cell}}$ and common $N_p$ the computational cost of each method is compared in Fig~\ref{fig:vacuum_complexity}.  The efficiency of all three methods was studied for the evaporation test case, where the physical domain was discretized along $x_2$ direction by $N_\mathrm{cell} = \{ 120, 140, 160 ,180,200\}$. The number of particles is set to $N_\mathrm{p} = N_\mathrm{cell}\times 2000$. As expected, both screened-Poisson and density expansion methods show a linear cost dependency on $N_{\textrm{cell}}$, whereas the direct methods cost scales quadratically. 

\section{Conclusion}
\label{sec:conclusion}

\noindent While an accurate description of dense flows including phase transition phenomena requires incorporating the long-range interactions, efficient simulations based on the Enskog-Vlasov kinetics remain a challenge. In particular, a brute-force computation of the Vlasov integral may arise to the complexity of the N-body problem. Based on the fact that these long-range interactions typically decay sharply for a neutral species, e.g. $1/r^6$ in the case of Sutherland's potential, a simple Taylor expansion of the number density can provide a computationally faster alternative. However, the underlying assumption behind the expansion relies on vanishing of the high order derivatives of the number density; an assumption which may not hold for relevant flow scenarios. \\ \ \\
\noindent In this study, a novel transformation is devised by approximating the attractive tail of the molecular potential, with Green's function of the screened-Poisson equation. Therefore, the corresponding Vlasov integral is translated to an elliptic equation which then can be solved by available efficient Poisson solvers. Note that in this framework, the Poisson equation arising from Coulomb interactions present in plasmas can be derived asymptotically. \\ \ \\
\noindent For validations and efficiency comparisons, two test cases were studied.  First, the planar Couette flow was simulated in a dense transitional regime. Next, evaporation of liquid argon below its critical temperature was considered. For all simulations, ESMC was adopted for computing short-range encounters, while the attractive forces were computed using the direct method, density expansion and the introduced screened-Poisson equation. Moreover, MD simulations were performed for further references.
Overall, a good agreement in terms of the accuracy is found between the above-mentioned methods. It is shown that the  screened-Poisson PDE can lead to more efficient computations with respect to the direct approach and yet more accurate predictions in comparison to the density expansion technique. For future studies, we plan to provide stochastic particle methods in order to compute the solution of the screened-Poisson PDE, through its associated random walk trajectories \cite{oksendalstochastic}. We anticipate that this development can lead to an all particle solution algorithm for the whole range of number densities. 
{\color{referee1}
 Moreover, we will pursue a detailed study on the accuracy of the screened-Poisson modelling approach in predicting multi-phase flow characteristics, such as the surface tension, interface thickness and evaporation/condensation rates of the droplets.
 }
\section*{Acknowledgements}
\noindent The authors gratefully acknowledge supports of Manuel Torrilhon throughout this study.
Furthermore, the authors appreciate the stimulating discussions with Patrick Jenny and Aldo Frezzotti. We would also like to thank the anonymous reviewers for their valuable comments and suggestions which significantly improved the quality of the paper. Hossein Gorji acknowledges the funding provided by Swiss National Science Foundation under the grant number 174060.

 \appendix
\bibliographystyle{elsarticle-num} 
  \bibliography{My_Collection}





\end{document}